\documentclass[12pt]{article}
\usepackage{amssymb}
\usepackage{amsmath}

\begin{document}

\begin{center}
\bigskip  \textit{Quantum theory: Reconsideration of Foundation-2}, editor
A. Yu. Khrennikov, (Vaxjo University Press, 2004), p. 315-322.

\bigskip {\large \textbf{How to Complete the Quantum-Mechanical Description?}%
}

\medskip Timur~F.~Kamalov

Physics Department, Moscow State Opened University,

ul. P. Korchagina, 22, Moscow 107996, Russia

Tel./fax 7-095-2821444

\begin{tabular}{ll}
E-mail: & qubit@mail.ru \\
& ykamalov@rambler.ru%
\end{tabular}
\end{center}

If the statement by Einstein, Podolsky and Rosen on incompleteness of
Quantum-Mechanical description of nature is correct, then we can regard
Quantum Mechanics as a Method of Indirect Computation. The problem is,
whether the theory is incomplete or the nature itself does not allow
complete description? And if the first option is correct, how is it possible
to complete the Quantum-Mechanical description? Here we try to complement
de-Broglie's idea on wave-pilot the stochastic gravitation gives origin to.
We assume that de-Broglie's wave-pilots are gravitational stochastic ones,
and we shall regard micro-objects as test classical particles being subject
to the influence of de-Broglie's waves stochastic gravitation.

Pacs 03.65*


\begin{center}
\bigskip

1. Introduction.
\end{center}

\bigskip

The Quantum Theory exists for many decades. But is everything OK with it
completeness[1]? To our opinion, it is not just so. The incompleteness of
Quantum-Mechanical description gives rise to various paradoxes, such as
Einstein-Podolsky-Rosen (EPR) one, the paradox of the Schroedinger's cat,
the Paradox of Quantum Nonlocality and Paradox of the Quantum Teleportation.
In this study we shall call the phenomena of quantum nonlocal behavior and
teleportation of the quantum states as paradoxes because they follow from
Stochastic Gravitation Model of Quantum Mechanic. It can be easily seen that
these are paradoxes, and indeed they are brought about by the drawbacks in
the Quantum Theory rather than being actual properties of nature. This is
due to the fact that time in Quantum Theory plays the role not conforming to
the physical reality. In particular, the Quantum Theory employs the concept
of Hilbert Space, in which time acts as a parameter. Henceforth, this
parameter (i.e. the time) may be the same in different points of the Hilbert
Space. This property of time in the Hilbert Space brings about the effects
of simultaneous quantum states of microobjects at different space points (or
transfer of the state from one Hilbert Space point to another with
velocities exceeding the velocity of light). These effects of the Quantum
Theory that are apparently real we call here the Quantum Nonlocality
Paradox. The Paradox of Quantum Teleportation is a sort of Quantum
Nonlocality Paradox. These paradoxes do not exist in the Classical Physics
and in the Stochastic Gravitation Model of the Quantum Mechanic, and General
Relativity Theory (employing the 4-dimensional space), in which different
points of time-space correspond to different values of the time.

And another question is whether the quantum-mechanical wave-function
interpretation of micro-objects is complete?

Let us consider the electron diffraction experiment on a set of the slits.
We shall consider this electron interference for the case when electrons
pass through the slits one-by-one with a small time gap between them. To
describe the observable pattern, we must solve the following dilemma: Either
each electron passes simultaneously through several slits, which seems
impossible from the classical physics viewpoint, or each of the
wave-electrons is coherent to others, which seems more correct and natural.
These wave-electrons must be coherent if the difference in their amplitudes
and phases is rather small and almost constant in time. Otherwise, the
interference pattern would be smeared due to varying amplitude and phase
difference.

Then, the question arises, why the wave functions within the
Quantum-Mechanical Description of different electrons are coherent. It is
perhaps more strange than the Quantum-Mechanical electron wave/particle
dualism. Postulating Quantum-Mechanical wave properties to be possessed by
each electron would not suffice, and to explain the interference pattern we
must complement the description with coherence of electron waves. This is
the additional requirement to account for electron interference. We can call
it the Phenomena of Quantum Coherence.

There exists a simple way to tell which of the slits has the electron passed
through. It is to leave open only a single slit. We can open any slit,
either the first or the second one, but we must leave open only one slit.
Maybe, it is possible to choose the time of opening and closing the slits so
that we see the interference pattern.

Now, let us consider the history of the Quantum Mechanical Description
incompleteness. In the EPR effect [1] two particles, $P$ and $Q$, interact
at the initial moment and then scatter in opposite directions. Let the first

of them be described by the wave function $\psi _{P}$ , the other by $\psi
_{Q}$. The system of the two particles $P$ and $Q$ is described by the wave
function $\psi _{PQ}$. For each of particles there are two distinct
descriptions depending on whether we take into account presence of the
second particle or not.

Where could the dependence of the object $P$ on the object $Q$ and vice
versa originate from, these objects $P$ and $Q$ being considered as distant
and non-interacting? The authors EPR came to the conclusion on
incompleteness of the quantum-mechanical description. To solve this
contradiction, an idea has been put forward in [1] on existence of hidden
variables that would make it possible to consistently interpret the results
of the experiments without altering the mathematical apparatus of quantum
mechanics.

Later, it has been proved by von Neumann [2] that quantum-mechanical
axiomatic does not allow introduction of hidden variables. It is, however,
important that the argument presented in [2] would not hold valid in certain
cases, e.g., for pairwise observable microobjects (for Hilbert space with
pairwise commutable operators) [4]. In 1964, J. S. Bell [5] has formulated
the experimental criterion enabling to decide, within the framework of the
problem statement [1], on the existence of the local hidden variables. The
essence of the experiment proposed by Bell is as follows. Let us consider
the following experimental scheme. Let there be two photons that can have
orthogonal polarizations $A$ and $B$ or $A^{\prime }$ and $B^{\prime }$,
respectively. Let us denote the probability of observation of the pair of
photons with polarizations $A$ and $B$ as $\psi _{AB}^{2}$. Bell has
introduced the quantity $\left| \left\langle S\right\rangle \right| =\frac{1%
}{2}\left| \psi _{AB}^{2}+\psi _{A^{^{\prime }}B}^{2}+\psi
_{AB^{^{\prime }}}^{2}-\psi _{A^{^{\prime }}B^{^{\prime
}}}^{2}\right| $, called the Bell's observable; it has been shown
that if the local hidden variables do exist, then $\left|
\left\langle S\right\rangle \right| \leq 1$. The possibility of
experimental verification of actual existence of local hidden
variables has been demonstrated in [5]. The above inequalities are
called Bell's inequalities. The series of experiments has shown
that there does not exist any experimental evidence of existence
of local hidden variables as yet, and the existing theories
comprising hidden variables are indistinguishable experimentally.
Because very interesting the contextualist viewpoint to the
probabilistic foundation of the quantum mechanics [6].

Further, considering the wave-pilot concept of de Broglie, we have to
complement it with the statement of these wave-pilots having to possess the
stochastic character in the space. These waves must generate mechanical
fluctuations of classical test particles; then, these particles can be
considered ``smeared'' in space and should be described by the
quantum-mechanical wave functions.

Einstein [2] has noted that it is impossible to extend geometrical
interpretation to the submolecular (sizes smaller than a molecule) scale, as
this would be as erroneous as to speak of a particle temperature for an
individual molecular-scale particle.

\begin{center}
2. Microobjects in the Curved Pro-Hilbert Space.
\end{center}

Recent years a very fascinating idea to put QM into geometric language
attracts the attention of many physicists. The starting point for such an
approach is the projective interpretation of the Hilbert space $\mathcal{H}$
as the space of rays. To illustrate the main idea it is convenient to
decompose the Hermitian inner product $\langle \cdot |\cdot \rangle $ in $%
\mathcal{H}$ into real and imaginary parts by putting for the two $L_{2}$%
--vectors $|\psi _{1}\rangle =u_{1}+\imath v_{1}$ and $|\psi _{2}\rangle
=u_{2}+\imath v_{2}$:

\begin{center}
$\langle \psi _{1}|\psi _{2}\rangle =G\,(\psi _{1},\psi _{2})-\imath \Omega
\,(\psi _{1},\psi _{2}),$
\end{center}

where $G$ is a Riemannian inner product on Projective Hilbert space or in
further description Pro-Hilbert space $\mathcal{H}$ and $\Omega $ is a
symplectic form, that is

\begin{center}
$G\,(\psi _{1},\psi _{2})=(u_{1},u_{2})+(v_{1},v_{2});\quad \Omega \,(\psi
_{1},\psi _{2})=(v_{1},u_{2})-(u_{1},v_{2}),$
\end{center}

with $(\cdot ,\cdot )$ denoting standard $L_{2}$ inner product. The
symplectic form $\Omega $ can acquire its dynamical content if one uses the
special stochastic representation of QM.

Let us consider two classic particles in random gravitational
fields or waves (in the relict gravitational background for
example) with number $j=1,2,3,...N$. In General Relativity Theory
the interval in this fields is

\begin{center}
$ds^{2}=\sum_{j=1}^{N}g_{\mu \nu }(j)dx^{\mu }dx^{\nu }=g_{\mu \nu
}^{0}dx^{\mu }dx^{\nu }$,
\end{center}

where the stochastic metric in the linear approach is

\begin{center}
$g_{\mu \nu }(j)=\eta _{\mu \nu }(j)+h_{\mu \nu }(j)$,
\end{center}

being $\eta _{\mu \nu }$ the Minkowsky metric, constituting the unity
diagonal matrix and $h_{\mu \nu }$ is perturbation of the metric. Here $%
g_{\mu \nu }^{0}=$ $\sum_{j=1}^{N}g_{\mu \nu }(j)$ we call a metric of the
Stochastic Curved Space. Hereinafter, the indices $\mu ,\nu ,\gamma ,m,n$
acquire values 0, 1, 2, 3. Indices encountered twice imply summation
thereupon.

Hilbert Space is non-curved space. But in Projective Hilbert space which we
are call the Pro-Hilbert Space the wave function $\psi ^{\mu }$ play the
role of coordinates.

Accepting Rieman's definition of the interval,

\begin{center}
$ds^{2}=g_{\mu \nu }^{0}dx^{\mu }dx^{\nu }=g_{\mu \nu }^{0}\frac{dx^{\mu }}{%
d\psi ^{i}}\frac{dx^{\nu }}{d\psi ^{k}}d\psi ^{i}d\psi ^{k}=G_{ik}d\psi
^{i}d\psi ^{k}=G^{ik}d\psi _{i}d\psi _{k}$
\end{center}

where $x^{\mu }$ being coordinates in the Rieman's space, $\mu ,\nu =0,1,2,3$
and $i,k=1,2,3,...,N$, where in common case $N\rightarrow \infty $,

we are denote

\begin{center}
$g_{\mu \nu }^{0}\frac{dx^{\mu }}{d\psi ^{i}}\frac{dx^{\nu }}{d\psi ^{k}}%
=G_{ik}$,
\end{center}

where $\psi ^{\mu }$ is wave function of microobject and $G_{\mu \nu }$ is
metric in Pro-Hilbert Space.

We will have the definition of the probabilities in Pro-Hilbert Space, if we
shall normalize the equation by condition

\begin{center}
$\int G_{\mu \nu }d\psi ^{\mu }d\psi ^{\nu }=P\leq 1$,
\end{center}

where $P$ means the probability's function with maximum $P=1$. The
probability $P$ in the Pro-Hilbert Space is the scalar product of two
vectors $\psi _{i}$ and $\psi _{k}$ with metric $G^{ik}$

\begin{center}
$P=G^{ik}\psi _{i}\psi _{k}$.
\end{center}

This space in common case is the curved. Really, if $l=1,2,3,...,N$, where $%
l $ is the number of a point, the volume of such figure is determined by the
formula

\begin{center}
$V_{N}=\frac{1}{N!}\left|
\begin{array}{cccc}
\psi _{1}^{1} & \psi _{2}^{1} & ... & \psi _{N}^{1} \\
\psi _{1}^{2} & \psi _{2}^{2} & ... & \psi _{N}^{2} \\
. & . & . & . \\
\psi _{1}^{N} & \psi _{2}^{N} & ... & \psi _{n}^{N}%
\end{array}
\right| $
\end{center}

In common case this space is curved, because $V_{n}\neq 0$ means that this
space is non-Evclidian.

The Hilbert Space with Evclidian metric $H_{ik}$\ is particular case of the
Pro-Gilbert Space with non-Evclidian metric $G_{ik}$ . If we denote $\Pi
_{ik}$ as weak perturbation of the Evclidian metric than

\begin{center}
$G_{ik}=H_{ik}+$ $\Pi _{ik},$ $\Pi _{ik}<<H_{ik\text{.}}$
\end{center}

Let us consider now the experiment with interference of two electrons on two
slits. The electron interference experiment is the one in which it is
impossible to determine the electron trajectory. Any attempt to determine
the electron trajectory fails due to any infinitesimal affecting of an
electron with the purpose of determination of its trajectory would alter the
interference pattern. This is the first aspect. On the other hand,
interaction of classical and stochastic fields and waves in such experiments
is usually neglected. Such interactions must exist in compliance with the
existing provisions of classical physics, and in particular, of the General
Relativity Theory. Moreover, this is experimentally confirmed by the
pre-quantum classical physics, hence, they require verification of their
effect onto quantum micro-objects.

Let us review some provisions of the General Relativity Theory. We consider
the motion of electrons from the source $S$ to the screen through slits 1
and 2.

\begin{center}
$\left\langle x\mid s\right\rangle _{1}=G_{ik}(1)\left\langle x\mid
s\right\rangle _{ik}=H_{ik}\left\langle x\mid s\right\rangle _{ik}+\Pi
_{ik}(1)\left\langle x\mid s\right\rangle _{ik}$,

$\left\langle x\mid s\right\rangle _{2}=G_{ik}(2)\left\langle x\mid
s\right\rangle _{ik}=H_{ik}\left\langle x\mid s\right\rangle _{ik}+\Pi
_{ik}(2)\left\langle x\mid s\right\rangle _{ik}$,
\end{center}

where $G_{ik}(1)\neq G_{ik}(2)$. Due to the propagation difference between
the two trajectories in space and time, the interference pattern is
generated. In the stochastic curved space one needs not to know the electron
trajectory. The interference pattern emerges due to the difference in
metrics $G_{ik}(1)$ and $G_{ik}(2)$. Thus, we have separated the wave
function of the space from the particle, because it is the property of the
space but not particle in our model. In Stochastic Gravitation Model the
microobjects is the test classical particle in the stochastic gravitation
fields and waves.

Let us select harmonic coordinates (the condition of harmonicity of
coordinates mean selection of concomitant frame $\frac{\partial h_{n}^{m}}{%
\partial x^{m}}=\frac{1}{2}\frac{\partial h_{m}^{m}}{\partial x^{n}}$) and
let us take into consideration that $h_{\mu \nu }$ satisfies the
gravitational field equations

\begin{center}
$\square h_{mn}(j)=-16\pi GS_{mn}(j)$,
\end{center}

which follow from the General Theory of Relativity; here $S_{mn}$ is
energy-momentum tensor of gravitational field sources with d'Alemberian $%
\square$ and gravity constant $G$. Then, the solution shall acquire the form

\begin{center}
$h_{\mu \nu }(j)=e_{\mu \nu }(j)\exp (ik_{\gamma }(j)x^{\gamma })+e_{\mu \nu
}^{\ast }(j)\exp (ik_{\gamma }(j)x^{\gamma })$,
\end{center}

where the value $h_{\mu \nu }(j)$\ is called metric perturbation, $e_{\mu
\nu }(j)$\ polarization, and $k_{\gamma }(j)$\ is 4-dimensional wave vector.

We shall assume that this metric perturbation $h_{\mu \nu }(j)$ is
distributed in space with an unknown distribution function $\rho =\rho
(h_{\mu \nu })$. Relative oscillations $\ell $ of two particles in classic
gravitational fields are described in the General Theory of Relativity by
deviation equations, which we can write for the stochastic case as

\begin{center}
$\frac{D^{2}}{D\tau ^{2}}\ell ^{\mu }(j)+R_{\nu mn}^{\mu }(j)\ell ^{m}\frac{%
dx^{\nu }}{d\tau }\frac{dx^{n}}{d\tau }=F(j)$,
\end{center}

being $R_{\nu mn}^{\mu }(j)$ the gravitational field Riemann's tensor with
gravitational field number $j$ of the stohastic gravitational fields and $%
F(j)$ is the stochastic constant (for the non-stochastic case this constant
is zero $F(j)=0$).

Specifically, the deviation equations give the equations for two particles
oscillations

\begin{center}
$\overset{..}{\ell }^{1}+c^{2}R_{010}^{1}\ell ^{1}=0,\quad \omega =c\sqrt{%
R_{010}^{1}}$.
\end{center}

The solution of this equation has the form

\begin{center}
$\ell ^{1}(j)=\ell _{0}\exp (k_{a}x^{a}+i\omega (j)t)$,
\end{center}

being $a=1,2,3$. Each gravitational field or wave with index $j$ and
Riemann's tensor $R_{\nu mn}^{\mu }(j)$ shall be corresponding to the value $%
\ell ^{\mu }(j)$ with stochastically modulated phase $\Phi (j)=\omega (j)t$.
If we to sum the all fields, we can write $\Phi (t)=\omega (t)t$, where $t$
is the time coordinate.

The stochastic phase $\Phi =\Phi (t)$ accounts for the Phenomena of Quantum
Coherence and we can use the stochastic phase $\Phi =\Phi (t)$ to understand
the nature of quantum interference.

\begin{center}
3. Bell's Inequalities as Experimental Criteria.
\end{center}

We shall consider the physical model with the Stohastic Gravitational
Background [i.e. with the background of gravitational fields and waves].
This means that we assume existence of fluctuations in gravitational waves
and fields expressed mathematically by metric fluctuations.

Considering quantum micro-objects in the stochastic curved space, we shall
take into consideration the fact that the scalar product of two 4-vectors $%
A^{\mu }$ and $B^{\nu }$ equals $g_{\mu \nu }^{0}A^{\mu }B^{\nu }$, where
for weak gravitational fields one can use the value $h_{\mu \nu }$, which is
the solution of Einstein's equations for the case of weak gravitational
field in harmonic coordinates.

Correlation factor $M$\ of random variables $\lambda ^{i}$\ are projections
onto directions $A^{\nu }$\ and $B^{n}$\ defined by polarizers (all these
vectors being unit) is

\begin{equation*}
\left| M_{AB}\right| =\left| \left\langle AB\right\rangle \right| =\left|
\langle \lambda ^{i}A^{k}g_{ik}\lambda ^{m}B^{n}g_{mn}\rangle \right|
\end{equation*}

The deferential geometry gives

\begin{center}
$\cos \Phi =\frac{g_{ik}\lambda ^{i}A^{k}}{\sqrt{\lambda ^{i}\lambda _{i}}%
\sqrt{A^{k}A_{k}}}$,

$\cos (\Phi +\theta )=\frac{g_{mn}\lambda ^{m}B^{n}}{\sqrt{\lambda
^{m}\lambda _{m}}\sqrt{B^{n}B_{n}}}$.
\end{center}

Here $i,k,m,n$ possess 0,1,2,3; $\theta $ is angle between polarizers, then

\begin{center}
$\left| M_{AB}\right| =\left| \frac{1}{2\pi }\int_{0}^{2\pi }\rho (\Phi
)\cos \Phi \cos \left( \Phi +\theta \right) d\Phi \right| =\left| \frac{1}{2}%
\rho \cos \theta \right| $,
\end{center}

if $\rho (\Phi )=\rho =const$ is the distribution function of $\Phi $.

Finally, the real part of the correlation factor for $\rho =2$ is

\begin{center}
$\left| M_{AB}\right| =\left| \cos\theta\right| $.
\end{center}

Then, we obtain the maximum value the Bell's observable $S$ in Rieman's
space for $\theta =\frac{\pi }{4}$

\begin{center}
$\left| <S>\right| =\left| \frac{1}{2}[\langle M_{AB}\rangle+\left\langle
M_{A^{\prime}B}\right\rangle +\left\langle M_{AB^{\prime}}\right\rangle
-\left\langle M_{A^{\prime}B^{\prime}}\right\rangle ]\right| =$

$=\left| \frac{1}{2}[\cos(-\frac{\pi}{4})+\cos(\frac{\pi}{4})+\cos(\frac{\pi
}{4})-\cos(\frac{3\pi}{4})]\right| =\left| \sqrt{2}\right| $,
\end{center}

which agrees fairly with the experimental data. The Bell in equality in
Rieman's space shall take on the form ${\left| \left\langle S\right\rangle
\right| \leq\sqrt{2}}$.

Therefore, we have shown that the Classical Physics with the Stohastic
Gravitational Background gives the value of the Bell's observable matching
both the experimental data and the quantum mechanical value of the Bell's
observable. To sum it up, the description of microobjects by the classical
physics accounting for the effects brought about by the Gravitational
Background is equivalent to the Quantum-Mechanical descriptions, both
agreeing with the experimental data.

\begin{center}
4. Conclusion.
\end{center}

We are describe here the Stochastic Gravitation Model of Quantum Theory with
stochastic gravitation de-Broigle waves \ which have the classical
gravitational fields and waves (i.e. Gravity Background with random nature)
origin.

Complementing the wave functions with the requirement of the Stochastic
Geometrical Fluctuation (in other words, with the stochastic nature of
gravitational fields and waves) enables us to get a new interpretation of
Quantum Mechanics. To put it otherwise, complementing the Quantum-Mechanical
description with stochastic gravitational fields and waves yields another
approach to Quantum-Mechanical microobject description

\begin{center}
Asknowlegment.
\end{center}

I'm grateful to Professor Yu. P. Rybakov from Russian University of Peoples
Friendship, Department Theoretical Physics and to Professor K. S. Kabisov
from Moscow State Opened University, Physics Department for useful
discussions.

\end{document}